\documentclass[twocolumn, showpacs, longbibliography, prl, aps, amsmath, amssymb,superscriptaddress]{revtex4-2}
\usepackage{amsmath}
\usepackage{amsfonts}
\usepackage{mathtools, amssymb}
\usepackage{graphicx}
\usepackage[colorlinks,linkcolor=blue,citecolor=blue,urlcolor=blue]{hyperref}
\usepackage{float}
\usepackage[normalem]{ulem}
\usepackage[left,columnwise, running]{lineno}
\usepackage[dvipsnames]{xcolor}
\usepackage{comment}

\begin{document}

\title{Ghost State of Light}

\author{R. M. de Boer}
\affiliation {Center for Nanophotonics, AMOLF, Science Park 104, 1098 XG Amsterdam, The Netherlands}

\author{C. Toebes}
\affiliation {Adaptive Quantum Optics, MESA+ Institute, University of Twente, PO Box 217, 7500 AE Enschede, The Netherlands}

\author{Jan Kl\"ars}
\affiliation {Adaptive Quantum Optics, MESA+ Institute, University of Twente, PO Box 217, 7500 AE Enschede, The Netherlands}

\author{S. R. K. Rodriguez} 
\email{s.rodriguez@amolf.nl}
\affiliation {Center for Nanophotonics, AMOLF, Science Park 104, 1098 XG Amsterdam, The Netherlands}

\begin{abstract}
We report the observation of a long-lived non-stationary state of light in a single-mode optical cavity. The observed state is a ghost of a saddle-node bifurcation which creates a bottleneck in phase space. While such ghosts are known to exist, accessing them is challenging because it requires a mechanism that steers the relaxation pathway away from the true attractor and into the bottleneck where the ghost emerges. Here we identify such a mechanism, namely a nonlinear response with memory. Our experimental system leverages this mechanism, enabling us to observe ghost states with lifetimes exceeding the cavity photon lifetime by more than ten orders of magnitude, even in the presence of strong fluctuations. The ghost manifests as a plateau in the relaxation dynamics of the  cavity transmission, reminiscent of prethermalization. We show how the ghost lifetime depends on the memory time and the distance to the bifurcation, and we observe signatures of scaling in the distribution of ghost lifetimes at fixed driving conditions. Our work establishes minimal conditions for realizing parametrically long-lived non-stationary states.

\end{abstract} 
\date{\today}
\maketitle

The relaxation time to a stationary state (RTSS) is a fundamental property of dynamical systems. In systems with many interacting components, the RTSS can emerge from the collective dynamics and vastly exceed all microscopic timescales. A paradigmatic example is the divergence of the RTSS at a continuous phase transition, known as critical slowing down~\cite{goldenfeld2018}. It underlies nonequilibrium phenomena such as the proliferation of topological defects described by the Kibble--Zurek mechanism~\cite{zurek2005,damski2005,dziarmaga2010,navon2015, du2023, lee2024}. Slow or inhibited relaxation can also arise away from criticality through different mechanisms. For example, the interplay of quenched disorder and many-body interactions in certain quantum systems can prevent thermalization for asymptotically long times, giving rise to many-body localization~\cite{nandkishore2015, abanin2019, schreiber2015,luschen2017, morong2021}. Another example is prethermalization, in which approximate conservation laws generate a separation of timescales enabling relaxation to a long-lived quasi-equilibrium before the system thermalizes~\cite{berges2004, moeckel2008, essler2016, reimann2019, luitz2020}. 

Different mechanisms preventing relaxation were recognized since the early days of nonlinear dynamics~\cite{pomeau1980,strogatz1994}. In particular, near a saddle-node bifurcation (SNB) where a stable and an unstable fixed point merge, the annihilated fixed points can leave behind a region in phase space acting as a bottleneck. If the system's trajectory passes through that region, the RTSS diverges and the system remains in a ghost state~\cite{sardanyes2020,canela2022,koch2024,zheng2025}. Theoretical interest in ghost states has recently surged,~\cite{gimeno2018, sardanyes2020, canela2022, koch2024, fucho2025, abouelela2025, koch2025, su2026}, yet experimental realizations remain scarce and limited to systems described by a single real-valued degree of freedom~\cite{trickey1998, noh2025}. A challenge to their observation in higher-dimensional systems is that phase-space trajectories can bypass the bottleneck along the system's relaxation path and no ghost state emerges.

In this Letter we experimentally demonstrate the emergence of a long-lived ghost state in a laser-driven optical cavity with thermo-optical nonlinearity. The ghost manifests as a plateau in the relaxation dynamics of the intracavity intensity at constant driving, similar to the plateau observed in prethermalization. We show that driving a cavity near a SNB is necessary but not sufficient for the ghost to emerge. An additional mechanism is needed to guide the phase-space trajectory into the bottleneck where the ghost emerges. In our system, that mechanism is memory in the nonlinear response. It reshapes the effective potential confining the phase-space dynamics, leading the optical field into a region where the phase space velocity vanishes and the ghost state emerges. We analyze the lifetime of this state in the absence of fluctuations as the SNB is approached, and in the presence of fluctuations at a fixed distance from the SNB. In both cases we find scaling behavior. Remarkably, even under strong fluctuations, we observe light trapping in a non-stationary state for a duration vastly exceeding all microscopic time scales in the cavity.


\begin{figure}
	\includegraphics[width=1.0\columnwidth]{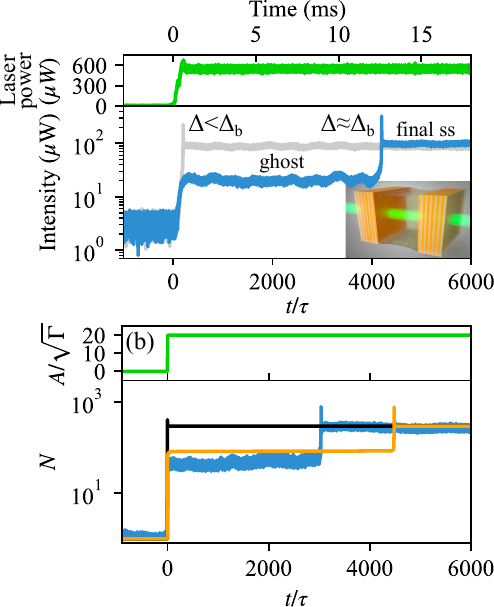}
	\caption{\label{fig1} (a) Experimental input (top, green curve) and transmitted (bottom) power of our optical cavity. Blue and gray curves correspond to a laser-cavity detuning close to and far away from the saddle-node bifurcation ($\Delta_b$), respectively. The inset in (a) is an illustration of the optical cavity under study. (b) Simulations corresponding to the experiments in (a). The green curve is the amplitude $A$ of the driving laser. Orange and blue curves are the intracavity intensity in the absence and presence of noise, respectively. The black curve corresponds to a cavity without memory $(\tau \rightarrow 0)$, in equal proximity to the saddle-node bifurcation as the orange and blue curves. Only in the presence of memory does the intensity shows a long-lived plateau before reaching the final steady state.} 
\end{figure}


The Fig.~\ref{fig1}(a) inset illustrates the system we study: a plano-concave Fabry-P\'erot microcavity driven by a continuous-wave laser and filled with olive oil. The oil's time-delayed thermo-optical nonlinearity effectively gives memory to the optical response~\cite{geng2020, peters2021, peters2022, braeckeveldt2024}. Technical details about our experimental setup are presented in Appendix A. Figure~\ref{fig1}(a) presents experimental results. We implemented a step in the laser power from zero to 550 $\mu$W, as shown by the green curve. We then monitored the transmitted intensity at two distinct laser wavelengths, judiciously detuned from the cavity resonance as explained ahead. The first detuning we consider places the cavity near a SNB ($\Delta_b$) after the power step. The resultant transmitted intensity is shown as a blue curve in Fig.~\ref{fig1}(a). Notice how the transmission quickly settles in a plateau after the power step. The intensity eventually rises from that plateau around 12 ms after the step, and the system relaxes to its unique steady state. The plateau's duration vastly exceeds both intrinsic timescales of our system. It surpasses the picosecond cavity photon lifetime (deduced from the resonance linewidth in the linear regime) by over 10 orders of magnitude, and the oil's microsecond thermal relaxation time (see supplemental material~\cite{supplemental}) by a factor of 4000. The second detuning we consider is further away from the same SNB after the power step. The resultant transmitted intensity, plotted as a grey curve in Fig.~\ref{fig1}(a), displays a quick relaxation without a plateau. Clearly, the driving conditions modify the relaxation dynamics. Next we introduce a model to explain the origin and duration of the plateau, which is the dynamical signature of a ghost state. The model we introduce has been validated through comparison with experimental observations in oil-filled cavities~\cite{geng2020,peters2021}.

In a frame rotating at the laser frequency $\omega$, the complex-valued intracavity light field $\alpha=\alpha_R+i\alpha_I$ satisfies:
\begin{align}
\label{eq1}
\dot{\alpha} &=\underbrace{\left(i\Delta  - \frac{\Gamma}{2} - iU\int_0^t ds\, K(t-s)N\right)\alpha + \sqrt{\kappa_L}A(t)}_{\vec{F}/\Gamma }  \nonumber\\
&\quad + \sqrt{2D_a}\xi(t) + i\delta(t)\alpha. 
\end{align}

\noindent $\Delta = \omega - \omega_0$ is the laser-cavity detuning, with $\omega_0$ the cavity resonance frequency.  $\Gamma = \gamma + \kappa_L + \kappa_R$ accounts for optical losses, including absorption at rate $\gamma$ and leakage through the left and right mirrors at rates $\kappa_{L,R}$. $U$ is the effective photon-photon interaction strength, which corresponds to the oil's refractive index change per unit temperature~\cite{geng2020}. The exact value of $U$ is irrelevant, since for $U\ll \Gamma$ (as in our experiments) the physics is entirely determined by the mean-field interaction strength $UN$ relative to $\Gamma$, where $N = |\alpha|^2$ is the intracavity intensity~\cite{carusotto2013}. $K(t) = \exp(-t/\tau)$ is a memory kernel, with $\tau$ the memory time given by the oil's thermal relaxation time~\cite{geng2020}. For $\tau \to 0$, memory effects vanish and the integral reduces to $UN$. In this limit, our time-delayed nonlinearity becomes the widely-studied Kerr nonlinearity~\cite{casteels2017, rodriguez2017, biondi2017, fink2018, Nigro25}. $A$ is the laser amplitude driving the cavity. All the above terms,  underbraced in Eq.~\eqref{eq1},  constitute the deterministic force $\vec{F}$ (divided by $\Gamma$) in our overdamped Langevin equation. The cavity is additionally driven by two stochastic forces. The first, $ \sqrt{2D_a}\xi(t)$,  accounts for Gaussian white noise due to both the driving laser and the cavity dissipation. The second,  $i\delta(t)\alpha$, includes an Ornstein-Uhlenbeck process accounting for a fluctuating laser-cavity detuning. The form of both stochastic forces is explained in Appendix B. To solve Eq.~\eqref{eq1} numerically, we perform a Markov embedding by defining $w=iU\int_0^t ds\, K(t-s)N$. Since the kernel is exponential, this leads to a separate equation for $w$: $\dot{w} = (UN - w)/\tau$.  Clearly,  $w$ is a ``hidden'' variable associated with the memory of our system. It is nonlinearly coupled to $\alpha$ via the term $UN$. $w$ can also be rigorously mapped to the temperature change in the thermo-optical medium filling our cavity~\cite{geng2020}.


\begin{figure}
	\includegraphics[width=1.0\columnwidth]{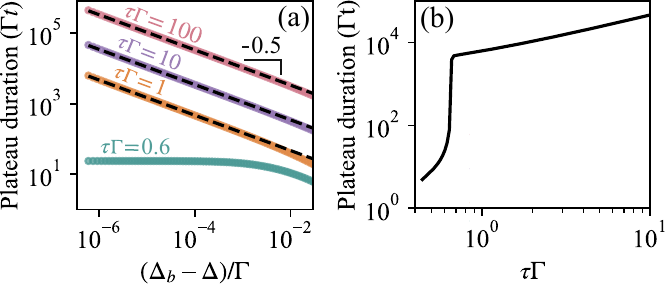}
	\caption{\label{fig2} (a) Plateau duration as a function of distance from the bifurcation for different memory times $\tau$. Dashed black lines are inverse square-root fits to the data. (b) Plateau duration as a function of $\tau\Gamma$ at $(\Delta-\Delta_b)/\Gamma = 6 \cdot 10^{-7}$.}
\end{figure}


For sufficiently large $A$ and $\Delta/\Gamma$, Equation~\eqref{eq1} predicts optical bistability: two stable steady states at a single driving condition~\cite{geng2020, peters2021}. The bistable region is enclosed by SNBs as shown in Appendix C. We performed numerical simulations (details in Appendix D) near such a SNB, in correspondence to the experiments in Fig.~\ref{fig1}(a). The results are presented in Fig.~\ref{fig1}(b). The green curve represents the laser amplitude $A(t)$. The orange curve is the resultant intracavity intensity in the absence of noise ($D_a, D_m = 0$). When noise is included, we obtain a trajectory like the one in blue. Both the orange and blue curves display a plateau like the one observed in experiments. The plateau is clearly robust to fluctuations, albeit its duration is reduced. The plateau disappears when the laser is further detuned from the SNB as shown ahead. A more insightful result from our model is obtained by setting $\tau = 0$ (corresponding to a memoryless Kerr-nonlinear cavity), while remaining in equal proximity to the SNB. The resultant trajectory, neglecting noise, is plotted as a black curve in Fig.~\ref{fig1}(b). The absence of a plateau in this case demonstrates that proximity to the SNB is insufficient for the ghost to emerge. A nonzero memory time is also needed. 

We investigated the effect of the memory time on the plateau duration (quantified as explained in supplemental material~\cite{supplemental}) through numerical simulations neglecting noise. The plateau duration is plotted as a function of the distance to the SNB for four distinct $\tau$ in Fig.~\ref{fig2}(a), and as a function of $\tau$ at fixed $\Delta$ in Fig.~\ref{fig2}(b). The distance to the SNB is controlled via the parameter $(\Delta_b-\Delta)/\Gamma$, with $\Delta_b$ the detuning at the SNB. The results show a qualitative change in behavior when the memory time becomes commensurate with the dissipation. For $\tau \gtrsim \Gamma$, Fig.~\ref{fig2}(a) shows that the plateau duration exhibits an inverse square-root scaling with $(\Delta_b-\Delta)/\Gamma$, which is the hallmark of a ghost state~\cite{strogatz1989, sardanyes2020}. Figure~\ref{fig2}(a) also shows a breakdown of the scaling law for a short memory time of $\tau \Gamma =0.6$. This positions the ghost, with its defining inverse square-root scaling, as a phenomenon emerging from the memory in our system. The transition around $\tau \sim \Gamma$ is explored in more detail in Fig.~\ref{fig2}(b). There we observe a sharp threshold in the plateau duration as $\tau$ increases, resulting in a linear scaling for $\tau\Gamma\gtrsim 1$. Next, we illustrate how the memory modifies the phase-space relaxation pathway.


\begin{figure}
	\includegraphics[width=1.0\columnwidth]{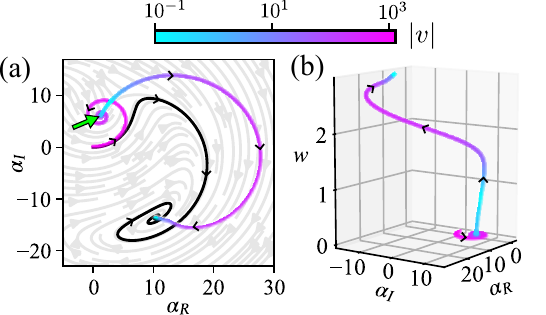}
	\caption{\label{fig3} (a) Relaxation trajectories in the complex $\alpha$ plane for the cavity with memory as a colored curve (with color representing instantaneous velocity) and for the memoryless cavity in black. The gray arrows represent the vector flow of the force $\vec{F}$ in Eq.~\eqref{eq1} with $\tau\to0$. (b) Relaxation trajectory for the cavity with memory in the three dimensional phase space of $\alpha_I$, $\alpha_R$, and $w$.} 
\end{figure}


Figure~\ref{fig3}(a) shows how the deterministic trajectories evolve in the complex $\alpha$ plane~\cite{delta}. For the cavity with memory, we plot the magnitude of the instantaneous velocity ($|v|=\sqrt{\dot\alpha_R^2+\dot\alpha_I^2}$) along the relaxation trajectory in color. For $\tau=0$, the trajectory is plotted in black. Both trajectories start at $\alpha_I=\alpha_R=0$, the steady state for $A=0$. When $A$ suddenly increases, the cavity with memory quickly evolves to the point indicated by the green arrow. Thereafter, the phase space velocity vanishes as the system passes through the bottleneck responsible for the ghost state. Interestingly, the bottleneck starts (green arrow tip) at the fixed point of a linear cavity ($U=0$). That point temporarily behaves as an apparent attractor because the nonlinearity is time-delayed, implying linear response immediately after the step. In contrast, the memoryless cavity bypasses the bottleneck since the nonlinearity is effective immediately after the step. Both cavities nonetheless reach the same unique steady state, as expected because the memory kernel does not affect the asymptotic behavior of our system~\cite{peters2021}. The results in Fig.~\ref{fig3}(a) illustrate how memory in the nonlinear response steers the trajectory into the bottleneck where the ghost emerges.

Figure~\ref{fig3}(b) shows the same relaxation trajectory for the cavity with memory as in Fig.~\ref{fig3}(a), but now in the entire phase space $(\alpha_R, \alpha_I, w)$. We can recognize three stages of evolution. First, $\alpha_R$ and $\alpha_I$ quickly relax within the plane $w=0$ towards the fixed point of a linear cavity (not the system under study). Second, $w$ increases slowly while $\alpha_R$ and $\alpha_I$ remain nearly constant. The ghost arises in this second stage, where the instantaneous velocity vanishes, and the intensity $N=\alpha_R^2 + \alpha_I^2$ is approximately constant while $w$ is evolving. This explains why the intensity plateau in Fig.~\ref{fig1} does not correspond to a stationary state: $w$ is evolving, but remains unobserved in an intensity measurement. Finally, in the third stage, all variables speed up on their approach to the unique steady state. This perspective reveals how the ghost is associated with a dynamical quasi-equilibrium in $\alpha$ that is far from true equilibrium.

\begin{figure}
	\includegraphics[width=0.9\columnwidth]{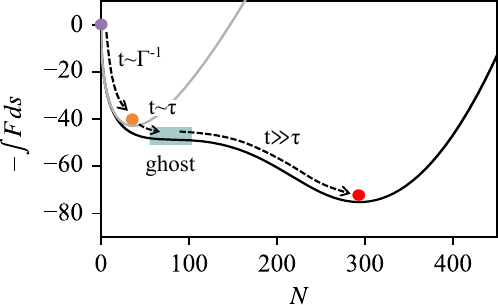}
	\caption{\label{fig4} Effective phase-space potentials at the start and end of our step-like protocol, plotted in gray and black, respectively. The flat region in the black potential is the bottleneck where the ghost state emerges.} 
\end{figure}





Next we illustrate why the phase space velocity of $\alpha$ vanishes in the ghost state. To this end, in Fig.~\ref{fig4} we calculate the effective potentials (see Appendix E for details) governing the relaxation of $\alpha$ immediately after the step and in the long time limit. These are plotted as gray and black curves, respectively. The effective potentials are one dimensional because we analyze the integral of $\vec{F}$ along the nullcline $\dot\alpha_I=0$. Despite this simplification, our analysis captures the essential physics because $\alpha_R$ is the only dynamical variable directly driven by the laser. 

Immediately after the power step, the nonlinearity has not yet developed. Hence, the corresponding effective potential (gray curve) is approximately harmonic. In contrast, once the nonlinearity has fully developed in the long-time limit, the corresponding effective potential (black curve) is anharmonic. We can now recognize the same three stages of evolution observed in Fig.~\ref{fig3}(b). Before the step, the system is at the location of the purple dot. Immediately after the step, the system relaxes to the bottom of the gray potential (orange dot) within the characteristic time $\Gamma^{-1}$; this is the first stage. Next, the nonlinearity builds up and the gray potential gradually transforms into the black one. Although the system always remains monostable, proximity to a SNB leads to a flat region in the black potential (green rectangle in Fig.~\ref{fig4}). As the velocity vanishes in that region, a bottleneck is formed. This is the second stage, where the ghost arises. Finally, in the third stage, the system escapes from the bottleneck and reaches the unique steady state indicated by the red dot in Fig.~\ref{fig4}.


We also studied the effect of noise on the ghost lifetime (plateau duration). To this end, we analyzed the distribution of ghost lifetimes obtained from 1000 independent experimental and numerical realizations. Experimental results are presented in Fig.~\ref{fig5}, and corresponding numerical results are in supplemental material~\cite{supplemental}. Within a limited range, the ghost lifetime distribution approximately follows a power law with exponent $-3/2$, plotted as a red line in Fig.~\ref{fig5}. This is similar to first passage time distributions of 1D Brownian motion~\cite{redner2023}. This follows from the fact that an integrated Ornstein–Uhlenbeck process, like $N$, behaves as a 1D Brownian motion~\cite{doob1942}.
Deviations from the $-3/2$ power law observed for long lifetimes are due to randomness in the initial detuning of our experiment. In supplemental material we demonstrate that, only if this randomness is included, the simulation reproduces the fat tail in the experimental distribution~\cite{supplemental}. Otherwise, the distribution follows the $-3/2$ power law.

\begin{figure}
	\includegraphics[width=1.0\columnwidth]{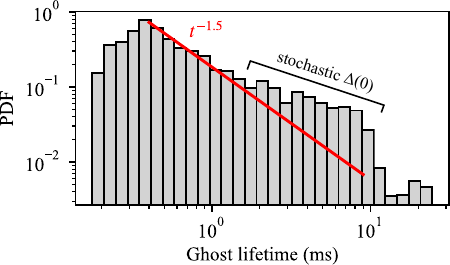}
	\caption{\label{fig5} Distribution of ghost lifetimes obtained from 1000 experimental trajectories. The red line is a power-law with exponent -3/2, which is the expected value (see supplemental material~\cite{supplemental}) in the absence of a stochastic initial condition in the detuning $\Delta$.}
\end{figure}

In conclusion, we have shown that a time-delayed nonlinearity can reshape relaxation pathways and steer trajectories into a phase-space bottleneck associated with a saddle-node bifurcation. This produces a long-lived ghost state, which we observed as a plateau in the transmission of our optical cavity. One perspective is to explore ghost states in the quantum regime. Recent studies of dissipative phase transitions in quantum systems have linked an ultralong RTSS to a vanishing Liouvillian spectral gap~\cite{casteels2017, rodriguez2017, fink2018, beaulieu2025, hadiseh2025}. However, that relaxation process is different from the one we discussed. It involves statistical mixing between mean-field steady states induced by quantum fluctuations. In contrast, the long RTSS associated with a ghost has a dynamical origin: it arises from a vanishing phase-space velocity, occurring even in the absence of fluctuations. The slow relaxation of a ghost state is not determined by barrier crossing between metastable configurations as in the case of dissipative phase transitions. Instead, it is a diffusion-limited process characterized by a slow wandering of the system through a nearly flat phase space region. Nonetheless, the connection between long-lived transients associated with metastable states and quantum effects such as entanglement~\cite{Munoz22, Bibak23, Torcal24} suggests that ghosts may open a new path to access and control quantum effects.

A second perspective is to realize ghost states in other systems that could enable unique functionalities. To this end, a cubic nonlinearity with memory is key. Two promising platforms fulfilling this condition are metasurfaces and cavity polaritons. Metasurfaces have recently demonstrated optical bistability arising from strong thermo-optical nonlinearities~\cite{Cotrufo24, King24, Barulin24}, indicating that ghost states are within reach. Ghosts in such spatially structured systems could enable the discovery of unconventional mechanisms to dynamically shape optical wavefronts. The other platform we envision, cavity polaritons, also exhibits bistability~\cite{carusotto2013, Sanvitto16, kavokin22} which is linked to a SNB. While polariton bistability is typically associated with an instantaneous nonlinearity due to polariton–polariton interactions, polaritons also interact with an exciton reservoir even under coherent driving~\cite{walker_dark_2017, stepanov_dispersion_2019, amelio_galilean_2020, Estrecho19, grudinina_dark_2022}. The equation of motion for the exciton reservoir is mathematically equivalent to our equation for $w$, suggesting that polariton ghosts are also within reach. Whether polariton ghosts can exhibit genuine quantum signatures is an interesting open question.



\section*{Acknowledgments}
\noindent This publication is part of the project ‘The cost of erasing an optical qubit’ with file number [NGF.1582.22.011] of the research programme NGF - Quantum Delta which is (partly) financed by the Dutch Research Council (NWO). We thank Long Him Cheung, Avishek Das, Johannes Feist, Antonio Fernandez Dominguez, Christopher Jarzysnki, Greg Stephens, and Pieter Rein ten Wolde, for stimulating discussions, as well as Niels Commandeur for technical support.


\begin{thebibliography}{65}
\bibitem [{del()}]{delta}%
  \BibitemOpen
  \href@noop {} {}\bibinfo {note} {$\Delta$ in Fig.~\ref{fig3} is slightly further away from the SNB than in Fig.~\ref{fig1}, but still within the red circle in Fig.~\ref{fig2}(a). $\Delta$ was changed for illustration purposes only. Otherwise, infinitesimally close to the SNB (as in Fig.~\ref{fig1}) the phase-space region of slow evolution is too small to be seen in Fig.~\ref{fig3}.}\BibitemShut {Stop}%
\end{thebibliography}

%

\section*{End Matter}

\emph{Appendix A: Experimental setup---} 
The cavity is made by a planar and a concave distributed Bragg reflector (DBR) mirror, each with a peak reflectance of 99.9\% at 530 nm. The DBRs of the planar and concave mirror consist of 11 alternating layers of SiO$_2$ and Ta$_2$O$_5$. The concave feature was made by direct laser writing \cite{vretenar2023, kurtscheid2020}. It has a diameter of 7.5 $\mu$m and a radius of curvature of 180 $\mu$m. We controlled the distance between the mirrors and their alignment by tuning screws in the corners of the sandwiched mirror mounts, which were separated by a stiff rubber material. The inter-mirror distance was fixed to $\sim$ 10 $\mu$m. The laser-cavity detuning was controlled with a tunable laser emitting in the 530-540 nm range. The step in the laser power was implemented with a mechanical chopper rotating at 20 Hz. Light was injected into the cavity and collected from it in transmission using 20x microscope objectives. The transmitted intensity was measured with an avalanche photodetector. The drop of olive oil responsible for the time-delayed nonlinearity was held in the cavity by capillary forces. We determined the thermal relaxation time of the oil from the overshoot it creates in the relaxation trajectory of the transmitted intensity, as previously explained~\cite{peters2021, peters2022EP}. Fig.~S1 shows a zoomed-in view of the overshoot in Fig.~\ref{fig1}(b). From the FWHM of this peak, we estimate a thermal relaxation time $\tau$ = 3 $\mu s$. This is a few microseconds shorter than in previous experiments~\cite{geng2020, peters2021}, possibly due to differences in mode volume and to the use of a different oil.

\emph{Appendix B: Stochastic forces---}
Here we explain the two stochastic forces driving $\alpha$ in Eq.~\eqref{eq1}.  The first stochastic force, $\sqrt{2D_a}\xi(t)$, contains a zero-mean delta-correlated complex Gaussian white noise with unit variance  $\xi(t) = \xi_R(t) + i \xi_I(t)$. The prefactor $\sqrt{2D_a}$ is the standard deviation of the stochastic force. This stochastic force accounts for both noise in the driving laser and dissipation noise, under the reasonable assumption that both noise sources are Gaussian and additive.  The  second stochastic force, $i\delta(t)\alpha$, accounts for a fluctuating laser-cavity detuning.  Here the noise is necessarily colored because the massive structures supporting our cavity mirrors suppress high-frequency fluctuations. We model this effect by modeling $\delta(t)$ as an Ornstein-Uhlenbeck process obeying the following equation of motion: $\dot{\delta}(t) = \delta(t)/\mu-\sqrt{2D_m/\mu} \zeta(t)$. $D_m$ and $\mu$ characterize the variance and correlation time of $\delta(t)$, respectively, while $\zeta(t)$ is a zero-mean delta-correlated Gaussian white noise with unit variance. $\delta(t)$ is multiplied by $\alpha$ in the same way that $\Delta$ is.

\begin{figure}
	\includegraphics[width=0.9\columnwidth]{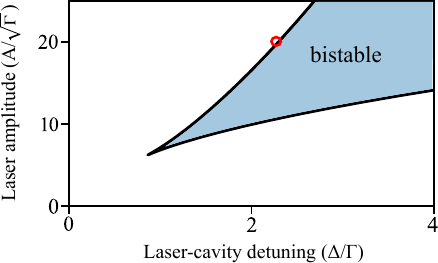}
	\caption{\label{figEM1} Region in parameter space where the cavity is bistable (blue), bounded by saddle-node bifurcations (black curves). The red circle indicates the driving conditions used in the main text after the step in laser power. Parameters: U=$\Gamma$/100, $\kappa$=$\Gamma$/2.}
\end{figure}

\emph{Appendix C: Bistable region and saddle-node bifurcations---}
Figure~\ref{figEM1} shows the region in parameter space where the cavity exhibits optical bistability. The blue area indicates the combinations of laser amplitude and laser-cavity detuning for which two stable solutions coexist. This bistable region is bounded by saddle-node bifurcations (SNBs), shown as black curves. The red circle marks the driving conditions used throughout the main text after the step in laser power. These parameters lie just outside the bistable region, such that the cavity is monostable but remains in close proximity to the SNB. Although only a single stable fixed point exists, the nearby saddle-node still imprints a bottleneck in phase space, giving rise to the long-lived plateau discussed in the main text.

\emph{Appendix D: Numerical Details---} 
We solved Eq.~\eqref{eq1} with the xSPDE MATLAB toolbox \cite{kiesewetter2016}. In all simulations we set $U=\Gamma/100$, $\kappa=\Gamma/2$, $\Delta=2.28\Gamma$ ($\Delta=2.299527\Gamma$ in Fig. \ref{fig1}(b)), and a laser amplitude $A=20\sqrt{\Gamma}$ after the power step. We used a numerical time step of $1/10\Gamma$ except for the simulations leading to the results in Fig.~\ref{fig2}, where we use a time step of $\tau/100$ or less if necessary to ensure accuracy. For the stochastic simulations in Fig.~\ref{fig1}, we set the noise variances to $D_a = 0.1\Gamma$ and $D_m = 0.05\Gamma$, and the noise correlation time to $\mu=500000/\Gamma$, which best reproduced the experimental observations. Due to computational limitations, we set a shorter thermal relaxation time $\tau$ in simulations than in experiments. Otherwise, to simulate a single experimental trajectory like the one in Fig.~\ref{fig1}, with a time step $1/10\Gamma \sim 0.1$ ps (the minimum necessary to ensure accuracy), we would need a dynamic range exceeding 11 orders of magnitude. To avoid such costly simulations, we set $\tau=100\Gamma^{-1}$. This value maintains the hierarchy of time scales present in the experiments, and thus enables us to qualitatively reproduce the results. Increasing it will simply extend the ghost lifetime in a linearly proportional way as shown in Fig.~\ref{fig2}(b).

\emph{Appendix E: Effective potentials---} 
Here we explain our approach to calculate the effective potentials in Fig.~\ref{fig4}. Those potentials govern the phase-space dynamics in the limits $t\to0$ and $t \to \infty$, where the nonlinear term in Eq.~\eqref{eq1} is completely absent and fully developed, respectively. Consequently, the dynamical contribution of the memory kernel to the potential can be neglected. 

We calculated effective one-dimensional potentials by integrating the deterministic force $\vec{F}$ in Eq.~\eqref{eq1} over a distance $s$ in the complex $\alpha$ plane, as done in Ref.~\onlinecite{peters2023}. The specific path that we choose for this integration is the nullcline $\dot{\alpha_I} = 0$. We previously rationalized this choice based on the fact that the dynamics along this path is limited to the real part of $\alpha$, $\alpha_R$, and this is the only dynamical variable directly driven by the laser. Next, we offer an additional explanation for the success of this approach based on the properties of $\vec{F}$ in the two-dimensional $\alpha$ plane.

\begin{figure}[H]
	\includegraphics[width=1.0\columnwidth]{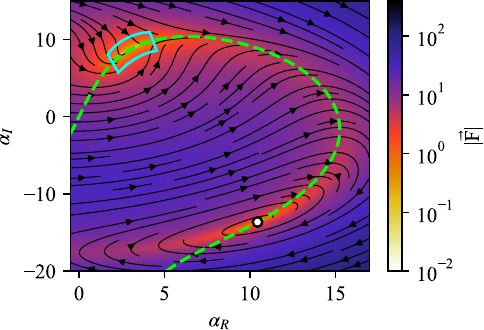}
	\caption{\label{figEM2} Force magnitude, in color, in the complex $\alpha$ plane at parameters close to the saddle-node bifurcation. Black arrows are the force vectors. The green dashed curve represents the $\dot{\alpha_I}$ nullcline. The white dot indicates the sole fixed point. Parameters: U=$\Gamma$/100, $\kappa$=$\Gamma$/2, $\Delta=2.299527\Gamma$, $A=20\sqrt{\Gamma}$.}
\end{figure}

Figure~\ref{figEM2} shows the magnitude of $\vec{F}$ in color, and its vector field as black arrows, for driving conditions close to the SNB where the ghost state emerges. The white dot indicates the location of the fixed point. This is the only attractor for the phase space dynamics. It is the only point where the magnitude of the force is strictly zero. Notice, however, that there is a yellow curved region where the magnitude of the force is very small and the phase space velocity is therefore reduced. Interestingly, that region is well traced by the dashed green curve we plotted on top, which is the nullcline $\dot{\alpha_I} = 0$. The overlap of this nullcline with the region of suppressed force explains the success of our approximate one-dimensional potential: the system tends to evolve towards regions of reduced force, which coincide with the nullcline $\dot{\alpha_I} = 0$. Notice, in particular, the region enclosed by the blue box, where the magnitude of the force is strongly suppressed. It is that region where the bottleneck responsible for the ghost state is formed.\\

\end{document}